\documentclass[12pt]{iopart}
\usepackage{amsfonts}
\usepackage{amssymb}
\usepackage{latexsym}
\usepackage{dsfont,eucal,mathrsfs}
\usepackage{epsfig}

\renewcommand{\mathcal}{\mathscr}

\newtheorem{teorema}{Theorem}

\newtheorem{coro}{Corollary}
\newtheorem{definition}{Definition}

\begin{document}

\title{Quantum States Allowing Minimum Uncertainty Product of $\phi$ and $L_z $}

\author{Tiago Pereira}\ead{tiagops@if.usp.br} 
\author{Domingos H. U. Marchetti}\ead{marchett@if.usp.br}
\address{Instituto de F\'{i}sica, Universidade de S\~ao Paulo,
C. P. 66318, 05315 S\~ao Paulo, SP, Brasil}

\begin{abstract}
We provide necessary and sufficient conditions for states to have an
arbitrarily small uncertainty product of the azimuthal angle $\phi $ and its
canonical moment $L_{z}$. We illustrate our results with analytical examples.
\end{abstract}

\section{Introduction}

The Newtonian determinism states that the present state of the universe
determines its future precisely. At the beginning of the past century the
advent of quantum mechanics exposed the determinism to great delusion. It
turned out that in the quantum world the uncertainty prevails. Heisenberg,
with his uncertainty principle, was the first to recognize the antagonism
between classical and quantum mechanics \cite{Heisenberg}. He notice that
for the position and its conjugate momentum the more concentrated the
distribution of the position, the more uniform is the distribution of the
momentum and vice-versa. The Heisenberg relation states that it is
impossible to predict, with arbitrary certainty, the outcomes of
measurements of two canonically conjugate observables.

The uncertainty relation was subsequently generalized by Robertson
\cite{Robertson}. The variance of an observable $A$ for a given state
$\psi $ is  
\[
\sigma _{A}^{2}=\langle A\psi ,A\psi \rangle -\left\vert \langle \psi ,A\psi
\rangle \right\vert ^{2}~, 
\]
and the Heisenberg-Robertson (HR) uncertainty relation, in its most well
known form, reads: 
\begin{equation}
\sigma _{A}\sigma _{B}\geq \frac{\hbar }{2}\left\vert \langle \psi
,i[A,B]\psi \rangle \right\vert ~,  \label{HR}
\end{equation}
where $[A,B]$ is the commutator of observables $A$ and $B$.

The uncertainty principle has been one of the most intricate points in
quantum mechanics \cite{EPR,Massar}. Besides its philosophical meaning it
plays a major role in experimental physics of atomic scale as, for example,
in the Bose-Einstein condensation \cite{BE}, and electrons jump at random
from one energy state which they could never reach except by fluctuations in
their energy. Another manifestation of the uncertainty principle in the
energy spectrum can be seen in the spectral linewidth that characterizes the
width of a spectral line \cite{Schawlow,Salerno}.

An old problem concerning the uncertainty principle and whether the
uncertainty relation (\ref{HR}) expresses it adequately appears if the
quantum system is described in terms of angle variables \cite{C}. When the
Cartesian coordinates $(x,y,z)$ are changed to spherical ones $(r,\theta
,\phi )$, equation (\ref{HR}) no longer provides a lower bound for the
product of uncertainty of the azimuthal angle operator $\phi $ and its
canonical conjugate momentum $L_{z}$ \cite{C,Jordan}. The trouble arises
since fluctuations on $\phi $ bigger than $2\pi $ do not have physical
meaning. Consequently, if $\psi $ is sufficiently localized in the Fourier
space, $\sigma _{L_{z}}$ is small $\sigma _{\phi }$ remains bounded and one
may have uncertainty product $\sigma _{\phi }\sigma _{L_{z}}$ smaller than
any given positive number. Recently this problem has attracted a great deal
of attention \cite{Jos,Tanimura,Folland,Franke,PeggNJP}.

The HR uncertainty relation for the angle and position has been criticized
on several grounds and other mathematical formulations of the uncertainty
principle have been proposed (see \cite{Jos,Folland,Kraus} for a
contextualization). Examples of such attempts include the entropic relations
relying on entropies instead of on the standard deviations of the
observables \cite{Deutsch,Maasan,Finkel,Birula}; by introducing a unitary
operator for phase $\phi $ \cite{Leblond,Holevo}; evaluating the commutator
for functions that just belong to the domains of the angle and angular
momentum operators \cite{PeggPRA,Gieres}; exchanging the angle with an
absolutely continuous periodic function \cite{Louisell}; and expressing the
lower bound as state dependent \cite{ Kraus,Judge}.

Despite of these alternatives, expressing the uncertainty principle for
angular operator by lower--bounding the product of the standard deviations
is widely used. In particular, experimental confirmation of the uncertainty
principle for the angular momentum and position has been carried out for
intelligent states (states that saturates the uncertainty relation for $\phi 
$ and $L_{z}$ observables) \cite{Franke}. Also recently, the relation
between these intelligent states and the constrained minimum uncertainty
product for the angular operator has shown to be important \cite{PeggNJP}.

Motivated by the state--dependence of standard measures of uncertainty and
the fact that some state features may be prepared or detected experimentally
we shall investigate the class of states that allows for an arbitrarily
small uncertainty product. For this, we introduce an one--parameter family
of states $\left\{ f_{\alpha }(\phi ),\alpha >0\right\} $, defined by the
Fourier coefficients of $f_{\alpha }(\phi )/A_{\alpha }$ \cite{CommentA} 
\begin{equation}
C_{n}(\alpha )=\frac{1}{2\pi A_{\alpha }}\int_{-\pi }^{\pi }e^{in\phi
}f_{\alpha }(\phi )d\phi \qquad ,\,n\in \mathbb{Z}  \label{Cn}
\end{equation}
with $A_{\alpha }$ fixed by the normalization $\int_{\pi }^{\pi }|f_{\alpha
}(\phi )|^{2}d\phi =1$ [see Eq. (\ref{A})].

In this paper, we provide necessary and sufficient conditions on these
families that allow for an arbitrarily small uncertainty product. We
demonstrate that arbitrarily small uncertainty product is attained if, and
only if, a single nonvanishing Fourier coefficient $C_{k}(\alpha )$ decays,
as a function of $\alpha $, slower than the others $C_{n}(\alpha )$
with $n\neq k$. Furthermore, we provide explicit examples of our
result. 

This paper is organized as follows: In Section \ref{Paradox} we discuss some
problems associated with the HR relation. Our hypotheses on the states are
given in Section \ref{SetUp}. Our main result concerning the states which
allow for an arbitrarily small uncertainty product is given in Section
\ref{Thm}. In Section \ref{Uncertainty} we deduce the equations for
$\sigma _{\phi }$ and $\sigma _{L_{z}}$. We provide examples of our result in
Section \ref{Exponentials} for the exponential decay and in Section
\ref{Poly} for the polynomial decay of the Fourier coefficients of the
states. In Section \ref{Cossine} we show that replacing $\phi $ by
$\sin \phi $ or $\cos \phi $ provides a good description of the HR
relation. Section \ref{Proof} contains a proof of our main
result. Finally, in Section \ref{Conclusion} we give our conclusions. 

\section{Pitfalls and Apparent Paradox}

\label{Paradox}

Let us start by introducing the operators $\phi $ and its canonical
conjugate $L_{z}$. The phase is introduced as the angular displacement of
the vector position: 
\[
\phi =\tan ^{-1}\left( \frac{y}{x}\right) . 
\]
The angle operator is usually defined as a multiplication operator either by
the variable $\phi $ or by \cite{Judge} 
\[
Y(\phi )=\left( \phi -\pi \right) \mbox{mod}~2\pi +\pi ~. 
\]
When $\phi $ is defined on the lift, that is, without the mod $2\pi $, it is
continuous but no longer periodic. Since $\phi $ and $\phi +2\pi $
correspond to the same physical situation, the mod $2\pi $ operation in the
range $\left[ -\pi ,\pi \right] $ is preferred. Here, we adopt $\phi $ as a
multiplication operator by $\phi $ acting on the space of $2\pi $--periodic
functions which is square integrable in the interval $\left[ -\pi ,\pi \right] $. 
For values in this range $\phi $ and $Y(\phi )$ do not differ
from each other.

The canonical momentum associated with $\phi $ is given by 
\begin{equation}
L_{z}=-i\hbar \left( x\frac{\partial }{\partial y}-y\frac{\partial }{
\partial x}\right) =-i\hbar \frac{\partial }{\partial \phi }.  \label{elizeh}
\end{equation}
Under the (false) assumption that the commutation relation 
\begin{equation}
\lbrack \phi ,L_{z}]=i\hbar  \label{comm}
\end{equation}
holds on the domain in which $L_{z}$ and $\phi $ are self--adjoint
operators, the HR uncertainty relation yields 
\begin{equation}
\sigma _{\phi }\sigma _{L_{z}}\geq \frac{\hbar }{2}.  \label{Uncer}
\end{equation}
The product of uncertainty, however, can be made smaller than $\hbar /2$ for
the majority of states \cite{Kraus,Franke,PeggNJP}.

Another apparent paradox that appears by na\"{\i}ve assumptions on the
domain of the operators involved is as follows. Let $\left\vert
lm\right\rangle $ denote the spherical harmonic functions. From
Eq. (\ref{comm}), we have  
\begin{equation}
\left\langle lm^{\prime }\right\vert [\phi ,L_{z}]\left\vert lm\right\rangle
=i\hbar \left\langle lm^{\prime }|lm\right\rangle ,~  \label{lm}
\end{equation}
and this leads to the (wrong) conclusion 
\[
\hbar (m-m^{\prime })\left\langle lm^{\prime }\right\vert \phi \left\vert
lm\right\rangle =i\hbar \delta _{mm^{\prime }}, 
\]
that $0=1$ if $m=m^{\prime }$. See Examples $\mathbf{5}$ and $\mathbf{6}$ of 
\cite{Gieres}.

Since the operator $\phi $ multiplies the wave function by a bounded real
number, it is Hermitian: $\langle \psi _{1},\phi \psi _{2}\rangle =\langle
\phi \psi _{1},\psi _{2}\rangle $, and self--adjoint operator in the Hilbert
space $\mathcal{H}$ of square integrable functions in $\left[ -\pi ,\pi \right] $. 
The operator $L_{z}$, on the other hand, is defined in a closed
domain $D(L_{z})$ of $\mathcal{H}$. It may be extended as a self--adjoint
operator if $D(L_{z})$ is the set of $2\pi $ --periodic absolutely
continuous functions $AC[-\pi ,\pi ]$ (see Section VIII.3 of \cite{simon}).
Now, the domain $D([\phi ,L_{z}])$ of the commutator $[\phi ,L_{z}]$ is
given by the functions $\psi \in AC[-\pi ,\pi ]$ such that $\psi (-\pi
)=\psi (\pi )=0$. As the eigenfunctions $\psi _{m}(\phi )=e^{im\phi }/\sqrt{
2\pi }$ of $L_{z}$ do not belong to $D([\phi ,L_{z}])$, the commutator
cannot acts over $\left\vert lm\right\rangle $ and equation (\ref{lm})
doesn't make sense. The apparent contradiction of (\ref{Uncer}) rests on the
same problem: the domain $D([\phi ,L_{z}])$ of functions in the
r.h.s. of (\ref{HR}) is smaller than the domain $D(L_{z})\cap D(\phi
)$ of the l.h.s. of (\ref{HR}) (see \cite{Gieres} for a detailed
discussion). 

An attempt to fix the domain problem in the uncertainty relation (\ref{Uncer}
) is to abandon the commutator and introduce a sesquilinear form \cite{Kraus,Gieres} 
defined in $D(L_{z})\cap D(\phi )$. The uncertainty relation
then reads 
\begin{eqnarray}
\sigma _{\phi }\sigma _{L_{z}} &\geq &\left\vert i\,\left\langle \phi \psi
,L_{z}\psi \right\rangle -i\left\langle L_{z}\psi ,\phi \psi \right\rangle
\right\vert  \nonumber \\
&=&\frac{\hbar }{2}\left\vert 1-2\pi \left\vert \psi (\pi )\right\vert
^{2}\right\vert  \label{Uncer1}
\end{eqnarray}
which is now state--dependent (see \cite{Gieres,Jos,PeggPRA}, for details).
Note that (\ref{Uncer1}) and (\ref{Uncer}) agree if $\psi \in D([\phi
,L_{z}])$, since a state $\psi $ in the domain of the commutator satisfies $\psi (\pi )=0$.

\section{Set Up}

\label{SetUp}

The ground of our result is the Fourier expansions of $f_{\alpha}(\phi )$: 
\begin{equation}
f_{\alpha}(\phi )=A_{\alpha} \sum_{n=-\infty }^{\infty
}C_{n}(\alpha)e^{in\phi },  \label{qualquer}
\end{equation}
where $C_{n}(\alpha)$ are the Fourier coefficients (frequency amplitudes) of 
$f_{\alpha}(\phi)/A_{\alpha}$, given by Eq. (\ref{Cn}), with $A_{\alpha}$
fixed by the normalization: 
\begin{eqnarray}
\langle f_{\alpha}(\phi ),f_{\alpha}(\phi ) \rangle &=&\int_{-\pi }^{\pi
}\left\vert f_{\alpha}(\phi )\right\vert ^{2}d\phi  \nonumber \\
&=&2\pi |A_{\alpha}|^{2}\sum_{n=-\infty }^{\infty }|C_{n}(\alpha)|^{2}=1.
\label{A}
\end{eqnarray}
\noindent

For notational simplicity, whenever we do not specify the sum we understand
the index running from $-\infty $ to $\infty $. Also, whenever there is no
risk of confusion, we shall omit the index $\alpha $ of the Fourier
coefficients $C_{n}$ and normalization constant $|A|^{2}$.

\smallskip

\noindent \textbf{Admissible Family:} Let $\mathcal{F}$ be an one parameter
family of periodic functions $f_{\alpha }$ with $(i)$ nontrivial variance,
that is, $\sigma _{\phi }^{2}\geq \inf_{\alpha }\sigma _{\phi }^{2}=\kappa
>0 $; and Fourier coefficients such that: $(ii)$ $\{nC_{n}(\alpha )\}\in
\ell _{2}$ uniformly in $\alpha $, that is, for every $\epsilon >0$ there is 
$N=N(\epsilon )$, independent of $\alpha $, such that $\sum_{n=j}^{m}n^{2}
|C_{n}(\alpha )|^{2}<\epsilon $ for all $m>j>N(\epsilon )$; $(iii)$ there is
an increasing sequence $(\alpha _{k})_{k\geq 1}$ such that $C_{n}(\alpha
_{j}) <C_{n}(\alpha _{k})$ if $j>k$. A family $\mathcal{F}$ is said to be
admissible if it satisfies $(i)$, $(ii)$ and $(iii)$.

Condition $(i)$ avoids a state $f_{\alpha }$ to be in a neighborhood of the
Dirac delta function $\delta (\phi )$. One expects $\left\vert f_{\alpha
}(\pi )\right\vert $ to be small for such states, so the bound given by Eq. (
\ref{Uncer1}) already prevents the uncertainty product to be close to $0$.
Condition $(ii)$ on uniformity is of technical nature and guarantees that
the limit of a sum equals to sum of the limits of a given sequence. It will
be used in Eqs. (\ref{Teste}) and (\ref{EqLim}). The last condition $(iii)$
is made here to give a relation of order inside the family, at least in
terms of subsequences, as $\alpha $ grows \cite{Comment1}.

\smallskip 

\noindent \textbf{Dominance Condition:} An admissible family $\mathcal{F}$
satisfies the dominance condition if within its one-parameter family of
Fourier Coefficients $\{C_{n}(\alpha )\}$ there is only one $C_{k}(\alpha
)\not=0$ $\forall \alpha $ such that \cite{CommentCi}  
\begin{equation}
\liminf_{\alpha \rightarrow \infty }\frac{C_{n}(\alpha )}{C_{k}(\alpha )}
=0,\,\,\forall n\not=k.  \label{CnC0}
\end{equation}

\section{Theorem on Arbitrarily Small Uncertainty Product}

\label{Thm}

Here we state our main results. We start by introducing the following

\begin{definition}
Let the standard deviations $\sigma _{\phi }(\alpha )$ and $\sigma
_{L_{z}}(\alpha )$ associated with a state $f_{\alpha }\in \mathcal{F}$ be
given by Eq. (\ref{variancia}) and (\ref{variancia1}), respectively. An
admissible family $\mathcal{F}$ is said to allow an arbitrarily small
uncertainty product if for every $\varepsilon >0$ there is an $\alpha ^{\ast
}\in (0,\infty )$ such that 
\[
\sigma _{\phi }(\alpha ^{\ast })\sigma _{L_{z}}(\alpha ^{\ast })<\varepsilon
.
\]
\end{definition}

Our main result is then stated as follows:

\begin{teorema}
\label{Theorem} An admissible family $\mathcal{F}$ allows an arbitrarily
small uncertainty product if, and only if, it satisfies the dominance
condition. 
\end{teorema}

From this theorem it follows:

\begin{coro}
Any state $f_{\alpha }(\phi )\in \mathcal{F}$ whose Fourier coefficients are
sufficiently localized in the Fourier space has uncertainty product smaller
than the least value predicted by the HR relation (\ref{Uncer}).
\end{coro}

It is worthy to note that our result does not depend on the decay of the
coefficients, but only on the relative decay with respect to $C_{k}$ as
stated in Eq. (\ref{CnC0}). We illustrate our findings for two different
decays. The proof of Theorem \ref{Theorem} is given in Section \ref{Proof}.

The consistency of Theorem \ref{Theorem} with the uncertainty relation (\ref
{Uncer1}) is as follows. The state $f_{\alpha }$ whose Fourier coefficients
satisfy the dominance condition (\ref{CnC0}) is such that $\left\vert
f_{\alpha }\right\vert ^{2}$ may be close to the uniform distribution for
some large $\alpha $ and this leads the r.h.s. of (\ref{Uncer1}) to be close
to $0$. Theorem \ref{Theorem} goes, however, beyond what the uncertainty
relation (\ref{Uncer1}) can predict. It follows, in particular, from the
prove of Theorem \ref{Theorem} that if the state $f_{\alpha }$ has two
\textquotedblleft dominant\textquotedblright\ Fourier coefficients, in the
sense of (\ref{CnC0}), the uncertainty product cannot be smaller than the
value predicted by relation (\ref{Uncer1}). In Section \ref{Poly}, we give
an examples of families of states of this type in which the uncertainty
product differs from the lower bound (\ref{Uncer1}) for all $\alpha $  (see
Fig. \ref{profile1}).

\section{Uncertainty Relations}

\label{Uncertainty}

In this section we give a formal derivation of the general formulas for the
deviations $\sigma _{\phi }$ and $\sigma _{L_{z}}$, assuming that Eq. (\ref
{qualquer}) holds. The deviation on the variable $\phi $ is given by: 
\begin{equation}
\sigma _{\phi }^{2}=\left\langle \phi ^{2}\right\rangle -\left\vert
\left\langle \phi \right\rangle \right\vert ^{2},  \label{variancia}
\end{equation}
and we start with the first term in the right-hand-side (r.h.s): 
\begin{eqnarray*}
\langle \phi ^{2}\rangle &=&\int_{-\pi }^{\pi }\phi ^{2}\left\vert f(\phi
)\right\vert ^{2}d\phi \\
&=&|A|^{2}\sum_{m,n}C_{m}^{\ast }C_{n}\int_{-\pi }^{\pi }e^{i(m-n)\phi }\phi
^{2}d\phi .
\end{eqnarray*}

Splitting the sum into $m\not=n$ and $m=n$, evaluating the integrals, and
using Eq. (\ref{A}) we have: 
\[
\langle \phi ^{2}\rangle =\frac{\pi ^{2}}{3}+4\pi |A|^{2}\xi 
\]
where 
\[
\xi =\sum_{m\not=n}C_{m}^{\ast }C_{n}\frac{(-1)^{(n-m)}}{(n-m)^{2}}. 
\]

For the second term in r.h.s. of (\ref{variancia}), we have 
\begin{eqnarray*}
\langle \phi \rangle &=&|A|^{2}\sum_{m,n}C_{m}^{\ast }C_{n}\int_{-\pi }^{\pi
}e^{i(n-m)\phi }\phi d\phi \\
&=&2\pi |A|^{2}\sum_{m\not=n}C_{m}^{\ast }C_{n}\frac{1}{i}\frac{(-1)^{n-m}}{
n-m}.
\end{eqnarray*}
Therefore the deviation is given by: 
\begin{eqnarray}
\sigma _{\phi }^{2} &=&\frac{\pi ^{2}}{3}+4\pi |A|^{2}\xi  \nonumber \\
&&-\left\vert 2\pi |A|^{2}\sum_{m\not=n}C_{m}^{\ast }C_{n}\frac{(-1)^{n-m}}{
n-m}\right\vert ^{2}.  \label{sigmaphi}
\end{eqnarray}

Next, we compute: 
\begin{equation}
\sigma _{L_{z}}^{2}=\langle L_{z}^{2}\rangle -\left\vert \langle
L_{z}\rangle \right\vert ^{2}~.  \label{variancia1}
\end{equation}
Using condition $(ii)$, we begin with 
\begin{equation}
\langle L_{z}^{2}\rangle =\langle L_{z}f,L_{z}f\rangle =|A|^{2}\hbar
^{2}\sum_{m,n}C_{m}^{\ast }C_{n}nm\int_{-\pi }^{\pi }e^{i(n-m)\phi }d\phi .
\label{Teste}
\end{equation}
The terms with $m\neq n$ vanish, while the terms with $n=m$ yield: 
\[
\langle L_{z}^{2}\rangle =2\pi |A|^{2}\hbar ^{2}\sum_{n}|C_{n}|^{2}n^{2}. 
\]

For the amount $\langle L_{z}\rangle =\langle f,L_{z}f\rangle $ we have
analogously 
\[
\left\vert \langle f,L_{z}f\rangle \right\vert ^{2}=4\pi ^{2}|A|^{4}\hbar
^{2}\left( \sum_{n}|C_{n}|^{2}n\right) ^{2}. 
\]
Thus, the deviation in $L_{z}$ is given by: 
\begin{equation}
\sigma _{L_{z}}^{2}=2\pi \hbar ^{2}|A|^{2}\sum_{n}|C_{n}|^{2}n^{2}-4\pi
^{2}\hbar ^{2}|A|^{4}\left( \sum_{n}|C_{n}|^{2}n\right) ^{2}.
\label{sigmaLz}
\end{equation}

\section{Fourier Coefficients with Exponential Decay\label{Exponentials}}

We restrict our attention to the case in which the frequency amplitudes $
C_{n}$ decay exponentially fast in $\left\vert n\right\vert $: 
\[
C_{n}=e^{-\alpha |n|}. 
\]

This and the next example capture most of the important features we wish to
emphasize. Note that, $C_{n}$ is a real even function of $n$: $C_{n}=C_{-n}$
and $C_{n}^{\ast }=C_{n}$. The sequence $\left\{ C_{n}(\alpha )\right\} $
satisfies hypotheses $(ii)$ and $(iii)$ but $f_{\alpha }(\phi )$ approaches
the Dirac delta function $\delta (\phi )$ when $\alpha $ tends to $0$: for
any piecewise continuous periodic function $\psi $, 
\[
f_{\alpha }\ast \psi (\phi )=\displaystyle\int_{-\pi }^{\pi }f_{\alpha
}(\phi -\zeta )\psi (\zeta )d\zeta \rightarrow (\psi (\phi +0)+\psi (\phi
-0))/2 
\]
and converges uniformly in any closed interval of continuity.

The sequence $\left\{ C_{n}(\alpha )\right\} $ satisfies, in addition, the
dominance condition Eq. (\ref{CnC0}) with $k=0$. As we shall see, the
uncertainty product can be arbitrarily small despite of the noncompliance of 
$(i)$.

From the properties of $C_{n}$ it follows that $\langle \phi \rangle =0$.
Note that the $1/(m-n)$ is odd, while the $C_{m}^{\ast }C_{n}(-1)^{n-m}$ is
even. As a result the product is odd, and a symmetric sum over an odd
function is zero. Therefore, we have 
\begin{equation}
\sigma _{\phi }^{2}=\frac{\pi ^{2}}{3}+2\frac{e^{2\alpha }-1}{e^{2\alpha }+1}
\xi (\alpha ),  \label{sigmaLzE}
\end{equation}
where 
\[
\xi (\alpha )=\sum_{m\not=n}e^{-\alpha |n|}e^{-\alpha |m|}\frac{(-1)^{n-m}}{
(n-m)^{2}}, 
\]
therein we have explicitly written the dependence of $\xi $ on $\alpha $. It
turns out that $(e^{2\alpha }-1)\xi (\alpha )/(e^{2\alpha }+1)$ is a
monotone increasing function of $\alpha \in \left( 0,\infty \right) $ and
the limit as $\alpha \rightarrow 0$ and $\alpha \rightarrow \infty $ always
exist. For the latter, we have 
\[
\lim_{\alpha \rightarrow \infty }\xi (\alpha )=0, 
\]
and an explicit computation shows that $\sigma _{\phi }^{2}=\displaystyle
\frac{\pi ^{2}}{ 3}(1+O(e^{-\alpha }))$ holds for large $\alpha $ (see \ref
{xi}). It thus follows that 
\begin{equation}
\lim_{\alpha \rightarrow \infty }\sigma _{\phi }^{2}=\frac{\pi ^{2}}{3},
\label{limIsigmaphi}
\end{equation}
is an upper bound for $\sigma _{\phi }^{2}$. Since $\sigma _{\phi }^{2}$
remains bounded for all values of $\alpha $, its physical significance is
assured. Note that $\sigma _{\phi }^{2}=\pi ^{2}/3$ is the deviation of a
uniform state $\psi (\phi )=1/\sqrt{2\pi }$, $\phi \in \left[ -\pi ,\pi 
\right] $.

The opposite situation yields: 
\[
2\lim_{\alpha \rightarrow 0}\frac{e^{2\alpha }-1}{e^{2\alpha }+1}\xi (\alpha
)=-\pi ^{2}/3. 
\]
In \ref{xi}, it is proved that, for $\alpha $ small enough, 
\begin{equation}
\sigma _{\phi }^{2}=\alpha ^{2}+O(\alpha ^{3})~,  \label{lim0sigmaphi}
\end{equation}
Hence, it yields 
\[
\lim_{\alpha \rightarrow 0}\sigma _{\phi }^{2}=0\ . 
\]

For the deviation $\sigma _{L_{z}}$ (since $C_{n}$ is even it implies $
\langle L_{z}\rangle =0$) we have 
\[
\sigma _{L_{z}}^{2}=2\hbar ^{2}\frac{e^{2\alpha }}{(e^{2\alpha }-1)^{2}}.
\]
In the limit $\alpha \rightarrow 0$ we obtain 
\begin{equation}
\sigma _{L_{z}}^{2}=\frac{\hbar ^{2}}{2\alpha ^{2}}(1+O(\alpha ))~,
\label{lim0sigmaL}
\end{equation}
and as $\alpha \rightarrow \infty $, we have 
\begin{equation}
\sigma _{L_{z}}^{2}=2\hbar ^{2}\frac{1}{e^{2\alpha }}\left( 1+O(e^{-2\alpha
})\right) .  \label{limIsigmaL}
\end{equation}
Hence, by Eq. (\ref{lim0sigmaphi},\ref{lim0sigmaL}), for $\alpha $ small
enough the uncertainty product 
\[
\sigma _{\phi }^{2}\sigma _{L_{z}}^{2}=\frac{\hbar ^{2}}{2}(1+O(\alpha )),
\]
asserts that the square of the uncertainty product reaches twice the
smallest predicted values by the HR relation (recall $f_{\alpha }(\phi )$
approaches $\delta (\phi )$ in this limit and it is not affected by the
boundary condition at $\pi $). For $\alpha $ large enough, by using
Eq. (\ref{limIsigmaL},\ref{limIsigmaphi}), we have  
\[
\sigma _{\phi }^{2}\sigma _{L_{z}}^{2}=\frac{2\pi ^{2}\hbar ^{2}}{3}\frac{1}{
e^{2\alpha }}(1+O(e^{-\alpha })),
\]
implying that the uncertainty product goes to zero exponentially fast
with $\alpha $. 

\begin{figure}[th]
\begin{center}
\epsfig{file=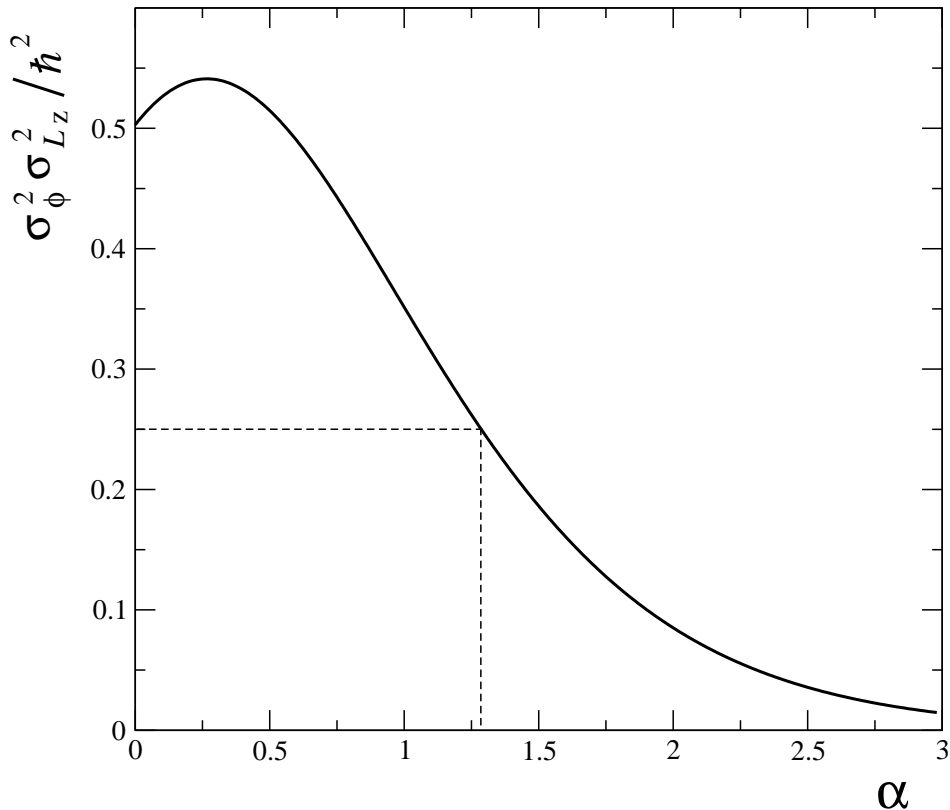, scale=0.65}
\end{center}
\caption{The profile of $\protect\sigma _{\protect\phi }^{2}\protect\sigma 
_{L_{z}}^{2}$ for a exponential decaying Fourier coefficients}
\label{profile}
\end{figure}

In Fig. \ref{profile} we depict the uncertainty product $\sigma _{\phi
}\sigma _{L_{z}}/\hbar $ as a function of $\alpha $. One can see that the
bound given by Eq. (\ref{Uncer}) holds only for $\alpha <1.29639$ (see the
dashed line).

\section{Polynomial Decay of Fourier Coefficients\label{Poly}}

The fact that the Fourier coefficients with exponential decay have an
arbitrarily small lower bound is not a privilege of this particular decay.
Any other decay which fulfills the hypotheses will also do so.

In our next example we want to illustrate that if the hypothesis of a unique 
$C_{k}$ in Eq. (\ref{CnC0}) is not fulfilled, the uncertainty product is
bounded from below as predicted by the HR uncertainty relation (\ref{Uncer}
). We consider a symmetric family of Fourier coefficients but we set $C_{0}$
to zero. As a consequence, there are two coefficients with the same decay as
a function of $\alpha $, and the dominance condition is no longer fulfilled
by the family. So, according to Theorem \ref{Theorem}, the uncertainty
product cannot be made arbitrarily small.

In the following, we shall consider 
\[
C_{n}=|n|^{-\alpha }\ ,\qquad n\neq 0 
\]
and $C_{0}=0$. If $\alpha \gg 1$ and $n\not=0$ the polynomial decay gives an
upper bound for the exponential decay. Note that in such limit $|n|^{-\alpha
}>\alpha ^{-|n|}$.

In this case, the normalization constant is given by 
\[
|A|^{2}=\frac{1}{2\pi \sum_{n}|n|^{-2\alpha }}. 
\]
The deviations now take the form 
\begin{eqnarray*}
\sigma _{\phi }^{2} &=&\frac{\pi ^{2}}{3}+\frac{1}{\sum_{n\geq 1}n^{-2\alpha
}}\sum_{m\not=n}|n|^{-\alpha }|m|^{{-\alpha }}\frac{(-1)^{(n-m)}}{(n-m)^{2}}
\\
\sigma _{L_{z}}^{2} &=&\frac{\hbar ^{2}}{\sum_{n\geq 1}n^{-2\alpha }}
\sum_{n\geq 1}n^{-2(\alpha -1)}
\end{eqnarray*}

In order to have $\sigma _{L_{z}}$ finite $\alpha $ must be bigger than $3/2$
, which guarantees that $|A|^{2}$ is larger than $0$. In the limit $\alpha
\rightarrow 3/2$ the deviation $\sigma _{L_{z}}$ diverges, while $\sigma
_{\phi }$ remains finite. The opposite situation yields: 
\begin{equation}
\lim_{\alpha \rightarrow \infty }\sigma _{\phi }^{2}\sigma _{L_{z}}^{2}=
\left(\frac{\pi ^{2}}{3} + \frac{1}{2} \right) \hbar ^{2} \approx 3.78986 \hbar ^{2} \label{sigmas}
\end{equation}
an uncertainty product larger than the least predicted value given by Eq. (\ref{Uncer}).

\begin{figure}[th]
\begin{center}
\epsfig{file=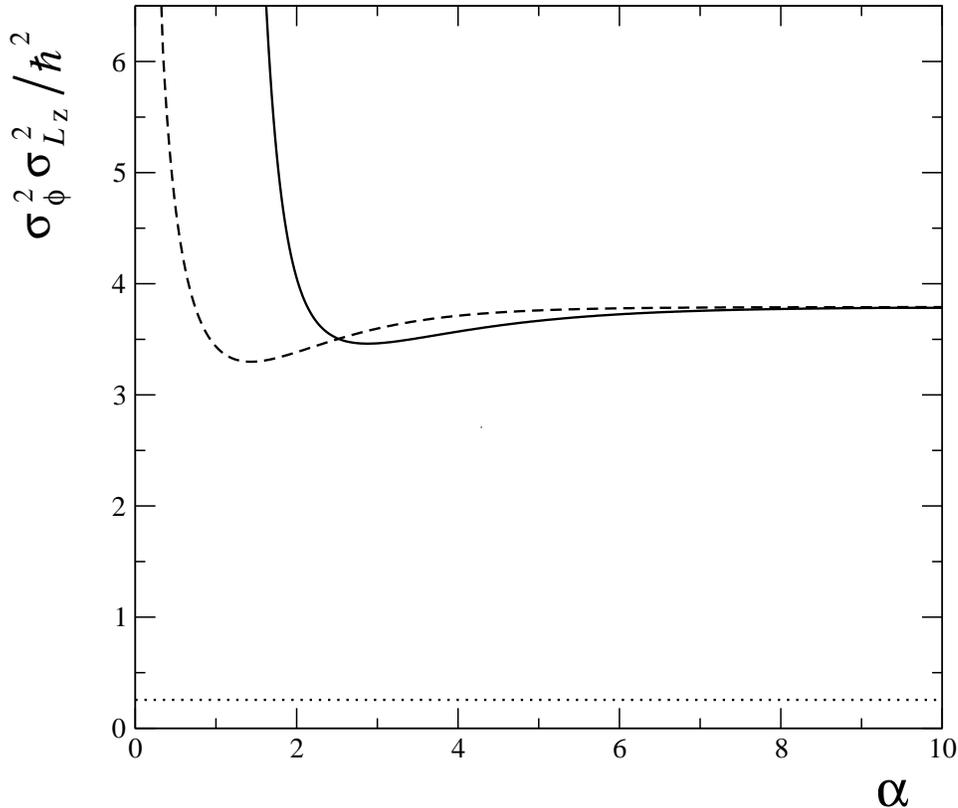, scale=0.65}
\end{center}
\caption{The profile of $\protect\sigma _{\protect\phi }^{2}\protect\sigma 
_{L_{z}}^{2}$ for polynomial (solid line) and  exponential (short
dashed line) decaying Fourier coefficients. Dashed line is the least
prediction of the HR uncertainty relation}
\label{profile1}
\end{figure}

Similar results hold for the exponential decay if we set $C_{0}=0$. The
profile of the uncertainty product for polynomial (solid line) and
exponential (short dashed line) decays, as a function of $\alpha $, are
shown in Figure \ref{profile1}.

\section{Replacing $\protect\phi $ by a Periodic Absolutely Continuous
Function\label{Cossine}}

As seen in Section \ref{Paradox}, the trouble with HR uncertainty relation (
\ref{Uncer}), with $A$ and $B$ replaced by angle operator $\phi $ and its
canonical conjugate momentum $L_{z}$, is not with the commutation relation (
\ref{comm}) but with the inequality 
\[
4\left\langle \phi \psi ,\phi \psi \right\rangle \left\langle L_{z}\psi
,L_{z}\psi \right\rangle \geq \left\langle \psi ,i\left[ \phi ,L_{z}\right]
\psi \right\rangle ^{2}~, 
\]
used to derive (\ref{Uncer}) from (\ref{comm}), which holds in a domain $
D\left( \left[ \phi ,L_{z}\right] \right) $ much smaller than the domain $
D\left( \phi \right) \cap D\left( L_{z}\right) $ of the left hand side.
Among the possibilities to overcome this problem, see \cite{Deutsch,
Maasan,Finkel,Birula,Holevo,PeggPRA,Gieres,Louisell,Judge}. Here we
illustrate the idea of replacing the operator $\phi $ by one periodic
operator that is absolutely continuous \cite{C,Louisell}. The basic idea is
to introduce the operators $\sin \phi $ and $\cos \phi $ which satisfy the
following commutation relation: 
\[
\lbrack \cos \phi ,L_{z}]=-i\hbar \sin \phi 
\]
and 
\[
\lbrack \sin \phi ,L_{z}]=i\hbar \cos \phi ~ 
\]
now defined in the domain $D\left( \sin \phi \right) \cap D\left(
L_{z}\right) =D\left( \cos \phi \right) \cap D\left( L_{z}\right) $.

In this way, we can compute the new uncertainty relations 
\begin{eqnarray}
\sigma_{ L_{z}}^2 \sigma_{ \sin \phi}^2 &\geq & \frac{\hbar ^{2}}{4}
\left\langle \cos \phi \right\rangle ^{2}  \label{uncersin} \\
\sigma_{ L_{z}}^2 \sigma_{ \cos \phi }^{2} &\geq & \frac{\hbar ^{2}}{4}
\left\langle \sin \phi \right\rangle ^{2}~.  \label{uncercos}
\end{eqnarray}
\noindent

Let us consider our previous example with the exponentially decaying
frequency amplitudes, now applying the new operators. The deviation 
\[
\sigma _{\cos \phi }^{2}=\langle \cos ^{2}\phi \rangle -\left\langle \cos
\phi \right\rangle ^{2}, 
\]
can be explicitly obtained. As a result we have 
\[
\sigma _{\cos \phi }^{2}=\frac{1}{2}\frac{e^{2\alpha }-e^{-2\alpha }+4}{
e^{2\alpha }+1}-4\frac{e^{2\alpha }}{(e^{2\alpha }+1)^{2}}. 
\]

For $\sin \phi $ the deviation is given by $\sigma _{\sin \phi }^{2}=\langle
\sin ^{2}\phi \rangle -\langle \sin \phi \rangle ^{2}$, and $\langle f,\sin
\phi f\rangle =0$. Thus after some manipulations we have 
\[
\sigma _{\sin \phi }^{2}=\frac{1}{2}\frac{e^{2\alpha }+e^{-2\alpha }-2}{
e^{2\alpha }+1}. 
\]

Note that for $\cos \phi $ we have the relation 
\[
\sigma _{L_{z}}\sigma _{\cos \phi }\geq 0, 
\]
since $\langle f,\sin \phi f\rangle =0$. This condition is always fulfilled.
The next relation we have to analyze is: 
\begin{equation}
\sigma _{L_{z}}\sigma _{\sin \phi }\geq \frac{\hbar ^{2}}{4}\langle \cos
\phi \rangle ^{2}.  \label{unS}
\end{equation}
Working the equations out, we have that Eq. (\ref{unS}) is equivalent to 
\[
e^{2\alpha }+e^{-2\alpha }-2\geq 0, 
\]
which is true for any $\alpha \geq 0$.

\section{Proof of the Main Results\label{Proof}}

For convenience, and pedagogic purposes, we consider the case of symmetric
Fourier coefficients $|C_{n}(\alpha )|=|C_{-n}(\alpha )|$. Theorem \ref{Theorem} 
states that the uncertainty product is arbitrarily small if, and only if, there is only 
one coefficient $C_{k}(\alpha )$ such that the rate $
C_{n}(\alpha )/C_{k}(\alpha )$ converges to zero as $\alpha $ grows
(dominance condition). For the symmetric case this coefficient must be 
\[
C_{0}(\alpha )=\frac{1}{2\pi A_{\alpha }}\int_{-\pi }^{\pi }f_{\alpha }(\phi
)d\phi 
\]
which is proportional to the spacial average of $f_{\alpha }$. $C_{0}(\alpha
)$ is the only possibility because otherwise it would always exist at least
two terms which, as a function of $\alpha $, decay slower than the other
coefficients. Thus, if a family of Fourier coefficient is symmetric and the
spacial average of the wave function is zero, our result implies in
particular that it is impossible to make $\sigma _{\phi }\sigma _{L_{z}}$ as
small as one wishes.

We start by showing that if the assumptions in Theorem \ref{Theorem} are
fulfilled then $\sigma _{\phi }\sigma _{L_{z}}$ is arbitrarily small. The
uncertainty of angular momentum is given by: 
\begin{equation}
\sigma _{L_{z}}^{2}=2\pi \hbar ^{2}|A|^{2}\sum_{n}|C_{n}(\alpha )|^{2}n^{2}~
\label{ss}
\end{equation}
with $A$ defined by (\ref{A}).

Given $\varepsilon >0$, we show that 
\begin{equation}
2\pi ^{3}\hbar ^{2}|A|^{2}\sum_{n}|C_{n}(\alpha )|^{2}n^{2}<\varepsilon 
\label{eqh}
\end{equation}
holds for some $\alpha =\alpha (\varepsilon )$. Introducing $|d_{n}(\alpha
)|^{2}=|C_{n}(\alpha )|^{2}/|C_{0}(\alpha )|^{2}$, Eq. (\ref{eqh}) is
equivalent to: 
\[
\frac{\pi ^{2}\hbar ^{2}}{\sum_{n}|d_{n}(\alpha )|^{2}}\sum_{n}|d_{n}(\alpha
)|^{2}n^{2}<\varepsilon .
\]
But since $\liminf_{\alpha \rightarrow \infty }|d_{n}(\alpha )|=0$ for all $
n\neq 0$, and the series $\sum_{n}|d_{n}(\alpha )|^{2}n^{2}$ is uniformly
convergent, by condition $(ii)$, we have 
\begin{equation}
\liminf_{\alpha \rightarrow \infty }\sum_{n}|d_{n}(\alpha
)|^{2}n^{2}=\sum_{n}\liminf_{\alpha \rightarrow \infty }|d_{n}(\alpha
)|^{2}n^{2}=0.  \label{EqLim}
\end{equation}
Note that $\sum_{n}|d_{n}(\alpha )|^{2}\geq 1$. Thus, by condition $(iii)$
for any $\varepsilon >0$ there is a $\alpha ^{\ast }$ such that 
\[
\frac{\pi ^{2}\hbar ^{2}}{\sum_{n}|d_{n}(\alpha ^{\ast })|^{2}}
\sum_{n}|d_{n}(\alpha ^{\ast })|^{2}n^{2}<\varepsilon .
\]

It follows from the definition of the deviation of ${\phi}$  
\[
\sigma_{\phi}^2 = \int_{-\pi}^{\pi} \phi^2 |f_{\alpha}(\phi)|^2 d\phi \le
\pi^2 \int_{-\pi}^{\pi} |f_{\alpha}(\phi)|^2 d\phi. 
\]
This implies $\sigma _{\phi }^{2} \leq \pi ^{2}$. Hence, it follows from (
\ref{ss}) and (\ref{eqh}) that 
\[
\sigma _{\phi }^{2}\sigma _{L_{z}}^{2}<\varepsilon , 
\]
and we finish the first part of the proof.

Next, we show the opposite implication. We want to show that outside our
hypothesis there exists $\varepsilon >0$ such that for all $\alpha \in
(0,\infty )$ 
\[
\sigma _{\phi }^{2}\sigma _{L_{z}}^{2}>\varepsilon \hbar ^{2} 
\]
and the uncertainty product cannot be made arbitrarily small.

Let $k\not=0$ be the smallest integer such that Eq. (\ref{CnC0}) holds, and
introduce $d_{n}(\alpha )=C_{n}(\alpha )/|C_{k}(\alpha )|$. Here, for sake
of simplicity, we assume that $k$ is unique, in the sense that only $|d_{-k}|
$ and $|d_{k}|$ are different from zero as $\alpha \rightarrow \infty $.

By $(i)$ we have $\sigma _{\phi }^{2}>\kappa $. Thus it suffices to
demonstrate that $\sigma _{L_{z}}^{2}$ is bounded away from zero. To this
end, we write 
\[
\sigma _{L_{z}}^{2}=\frac{\hbar ^{2}}{\sum_{n}|d_{n}(\alpha )|^{2}}
\sum_{n}|d_{n}(\alpha )|^{2}n^{2}. 
\]
We split the sum in the numerator and denominator as 
\[
\sum_{n}|d_{n}|^{2}=2+\sum_{\left\vert n\right\vert \not=k}|d_{n}|^{2} 
\]
and note that, by condition $(ii)$, there is $K<\infty $ independent of $
\alpha $ such that $\sum_{|n|\not=k}|d_{n}|^{2}\leq 2K$. Hence, 
\[
\sigma _{L_{z}}^{2}\geq \frac{\hbar ^{2}}{1+K}\left(
k^{2}+\sum_{n\not=k}|d_{n}|^{2}n^{2}\right) \geq \frac{\hbar ^{2}}{1+K} 
\]
in view of $\sum_{n\not=k}|d_{n}|^{2}n^{2}\geq 0$ and $k\geq 1$. The
uncertainty product can be bounded from below by 
\[
\sigma _{\phi }^{2}\sigma _{L_{z}}^{2}>\kappa \frac{\hbar ^{2}}{1+K} 
\]
Since $K$ does not depend on $\alpha $ and $\kappa >0$ is fixed, we can take 
$\varepsilon >0$ so that $\kappa /(1+K)>\varepsilon $, concluding 
\[
\sigma _{\phi }^{2}\sigma _{L_{z}}^{2}>\varepsilon \hbar ^{2}~. 
\]

Our result also holds for asymmetric Fourier coefficients. We do not
consider it here since the arguments are the same as for the symmetric case
with further technicalities.

\section{Conclusions\label{Conclusion}}

In conclusion, we have analyzed the uncertainty product for the azimuthal
angle $\phi $ and its canonical conjugate moment $L_{z}$. We have provided
necessary and sufficient conditions for a state to have an arbitrary small
uncertainty product. These conditions are related to the existence of a
Fourier coefficient of $f_{\alpha }$ which decays slower than the others
Fourier modes. More precisely, a state allows for an arbitrary small
uncertainty product if, and only if, there is only one coefficient $C_{k}(\alpha )$, 
such that $\lim \inf_{\alpha \rightarrow \infty}C_{n}(\alpha )/C_{k}(\alpha )=0$ 
(the dominance condition).

\bigskip 

\noindent \textbf{Acknowledgment.} We would like to thank Prof. W. F.
Wreszinski for important discussions concerning the problem. We also thank
F. Gieres and S. Tanimura for pointing out some useful references on the
topic. We are in debt to Dr. A. Veneziani for his careful and critical
reading of the manuscript. We are grateful to the peer referees whose
criticism and pertinent suggestions helped to improve the manuscript. T. P.
is supported by FAPESP, Project 07/04579-2.

\appendix

\section{Estimation of $\protect\sigma _{\protect\phi }^{2}$ for $\lim 
\protect\alpha \rightarrow 0$\label{xi}}

Proceeding the variable change $k=n-m$ in $\xi (\alpha )$ we have 
\[
\xi (\alpha )=\sum_{k\not=0}\frac{(-1)^{k}}{k^{2}}\sum_{n}e^{-\alpha
|n|}e^{-\alpha |n-k|}. 
\]
Due to the modulo we must split the above equation as follows: 
\begin{eqnarray*}
\xi (\alpha ) &=&\sum_{k\geq 1}\frac{(-1)^{k}}{k^{2}}\left[ \sum_{n\geq
k}e^{-\alpha n}e^{-\alpha n}e^{\alpha k}+\sum_{0\leq n<k}e^{-\alpha
n}e^{\alpha n}e^{-\alpha k}+\sum_{n<0}e^{\alpha n}e^{\alpha n}e^{-\alpha k}
\right] \\
&+& \sum_{k\leq -1}\frac{(-1)^{k}}{k^{2}}\left[ \sum_{n>0}e^{-\alpha
n}e^{-\alpha n}e^{\alpha k}+\sum_{k<n\leq 0}e^{\alpha n}e^{-\alpha
n}e^{\alpha k}+\sum_{n\leq k}e^{\alpha n}e^{\alpha n}e^{-\alpha k}\right] .
\end{eqnarray*}

This can also be written as: {\small 
\[
\xi (\alpha )=2\sum_{k\geq 1}\frac{(-1)^{k}}{k^{2}}\left[ \left(
\sum_{n>0}e^{-2\alpha n}+k\right) e^{-\alpha k}+\sum_{n\geq k}e^{-2\alpha
n}e^{\alpha k}\right] 
\]
} Noting that $\sum_{n\geq k}e^{-2\alpha n}=e^{2\alpha }e^{-2\alpha
k}/(e^{2\alpha }-1)$, then 
\[
\xi (\alpha )=2\sum_{k\geq 1}\frac{(-1)^{k}}{k^{2}}\left( \frac{e^{2\alpha
}+1}{e^{2\alpha }-1}+k\right) e^{-\alpha k} 
\]

Thus, we deviation takes the form: 
\begin{equation}
\sigma _{\phi }^{2}=\frac{\pi ^{2}}{3}+4\sum_{k\geq 1}\frac{(-1)^{k}}{k^{2}}
e^{-\alpha k}+4\frac{e^{2\alpha }-1}{e^{2\alpha }+1}\sum_{k\geq 1}\frac{
(-1)^{k}}{k}e^{-\alpha k}~.  \label{sigma2}
\end{equation}
\qquad

Introducing 
\begin{equation}
g(\alpha )=4\frac{e^{2\alpha }-1}{e^{2\alpha }+1}\sum_{k\geq 1}\frac{(-1)^{k}
}{k}e^{-\alpha k}~,  \label{g}
\end{equation}
in the limit of small $\alpha $ we have 
\[
\lim_{\alpha \rightarrow 0}\sigma _{\phi }^{2}=\frac{\pi ^{2}}{3}
+4\sum_{k\geq 1}\frac{(-1)^{k}}{k^{2}}\lim_{\alpha \rightarrow 0}e^{-\alpha
k}+\lim_{\alpha \rightarrow 0}f(\alpha ),
\]
which equals 
\[
\lim_{\alpha \rightarrow 0}\sigma _{\phi }^{2}=\lim_{\alpha \rightarrow
0}g(\alpha ),
\]
since $\displaystyle\sum_{k\geq 1}\displaystyle\frac{(-1)^{k}}{k^{2}}=-\pi
^{2}/12$. To estimate $g(\alpha )$, we note that 
\begin{eqnarray}
\sum_{k\geq 1}\frac{(-1)^{k}}{k}e^{-\alpha k} &=&\int_{\alpha }^{\infty
}\sum_{k\geq 1}(-1)^{k}e^{-\zeta k}d\zeta   \nonumber \\
&=&-\int_{\alpha }^{\infty }\frac{e^{-\zeta }}{1+e^{-\zeta }}d\zeta  
\nonumber \\
&=&-\ln \left( 1+e^{-\alpha }\right)   \label{sum}
\end{eqnarray}
since the series converges absolutely for $\alpha >0$ and the sum can be
performed before the integral. Thus, 
\[
g(\alpha )=-4\frac{e^{2\alpha }-1}{e^{2\alpha }+1}\ln \left( 1+e^{-\alpha
}\right) 
\]
The expansion in power of $\alpha \ll 1$ up to third order gives 
\begin{equation}
g(\alpha )=-4\alpha \ln 2+2\alpha ^{2}+O(\alpha ^{3}).  \label{gexp}
\end{equation}
Consequently, $\lim_{\alpha \rightarrow 0}g(\alpha )=0$ and 
\[
\lim_{\alpha \rightarrow 0}\sigma _{\phi }^{2}=0.
\]

Eq. (\ref{sigma2}) can be written in a closed form as 
\begin{equation}
\sigma _{\phi }^{2}=\frac{\pi ^{2}}{3}+4\mathrm{Li}_{2}(-e^{-\alpha
})+g(\alpha ),  \label{sigmali}
\end{equation}
where $\mathrm{Li}_{2}(z)=\sum_{k>0}z^{k}/k^{2}$ is the dilogarithm
function, whose series in power of $\alpha $ up to order $3$ is given by 
\begin{equation}
\mathrm{Li}_{2}(-e^{-\alpha })=\displaystyle\sum_{k\geq 1}\displaystyle\frac{
(-1)^{k}}{k^{2}}e^{-\alpha k}=\frac{-\pi ^{2}}{12}+\alpha \ln 2-\frac{\alpha
^{2}}{4}+O(\alpha ^{3})~.  \label{li2exp}
\end{equation}

Replacing Eqs. (\ref{gexp}) and (\ref{li2exp}) in Eq. (\ref{sigmali}) it
yields 
\[
\sigma _{\phi }^{2}=\alpha ^{2}+O\left( \alpha ^{3}\right) ,
\]
which dictates the behavior of the product $\sigma _{\phi }\sigma _{L_{z}}$
as $\alpha \rightarrow 0$, as can be seen in Fig. \ref{profile}.


\begin{thebibliography}{99}
\bibitem{CommentA} We write $f_{\alpha}(\phi)/A_{\alpha}$ for convenience,
so we can handle, for example, the Fourier coefficients and Eq. (\ref{qualquer}) 
easily.
\bibitem{Comment1} Alternatively, we could require the opposite situation 
$C_{n}(\alpha _{j})<C_{n}(\alpha _{k})$ if $k>j$ as $\alpha $ goes to zero,
or to a fixed $\alpha _{0}$.
\bibitem{CommentCi} The lim inf in Eq. (\ref{CnC0}) means that there is at
least one subsequence $\left( \alpha _{j}\right) _{j\geq 1}$ such that 
$\lim_{j\rightarrow \infty }C_{n}(\alpha _{j})/C_{k}(\alpha _{j})=0$ for all $
n\neq k$. No such subsequence exists if $\liminf_{\alpha \rightarrow \infty
}C_{n}(\alpha )/C_{k}(\alpha )>0$ for some $n\neq k$.
\end{thebibliography}
\end{document}